# Iron spin-reorientation transition in NdFeAsO

A. Marcinkova[1], T.C. Hansen[2], and J.W.G. Bos[3*]

*1. School of Chemistry and Centre for Science at Extreme Conditions, University of Edinburgh, West Mains Road, Edinburgh, EH9 3JJ, United Kingdom.*

*2. Institut Laue Langevin, 38042, Grenoble, France.*

*3. Institute of Chemical Sciences and Centre for Advanced Energy Storage and Recovery, Heriot-Watt University, Edinburgh, EH14 4AS, United Kingdom.*

* j.w.g.bos@hw.ac.uk

The low-temperature magnetic structure of NdFeAsO has been revisited using neutron powder diffraction and symmetry analysis using the Sarah representational analysis program. Four magnetic models with one magnetic variable for each of the Nd and Fe sublattices were tested. The best fit was obtained using a model with Fe moments pointing along the $c$-direction, and Nd moments along the $a$-direction. This signals a significant interplay between rare-earth and transition metal magnetism, which results in a spin-reorientation of the Fe sublattice upon ordering of the Nd moments. All models that fit the data well, including collinear models with more than one magnetic variable per sublattice, were found to have an Fe moment of 0.5 $\mu_B$ and a Nd moment of 0.9 $\mu_B$, demonstrating that the low-temperature Fe moment is not substantially enhanced compared to the spin-density wave (SDW) state.

High-temperature superconductivity and magnetism are intricately linked in the iron based high-$T_c$ superconductors [1-3]. The interplay between the $3d$ and $4f$ magnetism in 1111-type RFeAsO materials is therefore of much interest. The most prominent examples include CeFeAsO, where the Fe ordering induces a substantial magnetization on the Ce sublattice [4], and CeFePO, which is a heavy Fermion metal [5], whereas all other RFePO materials are low-temperature superconductors. Recently, strong coupling between the Sm and Fe magnetism in SmFeAsO was established from magnetic X-ray scattering [6]. In NdFeAsO, a substantial enhancement of the Fe moment upon Nd ordering was reported from neutron powder diffraction [7]. In fact, considerable uncertainty remains in the literature regarding the ordered magnetic structures of these materials. For example, several different low-temperature magnetic structures have been reported for PrFeAsO [8-10], and the magnitude of the Fe moment has proved controversial with neutron diffraction suggesting different values for different R [3]. Spectroscopic techniques, in contrast, consistently indicate ordered moments of ~0.4 $\mu_B$ in the SDW state [4, 11, 12]. The main cause of this uncertainty is the difficulty in solving magnetic structures in high-symmetry materials with multiple sublattices that contribute to the same Bragg reflections. Representational analysis of the symmetry allowed magnetic structures is very insightful in these situations [13], and has recently been used to determine the magnetic structures of a variety of RTMAsO (TM = Mn, Fe, Co) materials [4, 6, 14-17].

There is currently no symmetry assisted analysis of the low-temperature (Nd + Fe) magnetic structure of NdFeAsO in the literature. This manuscript provides this missing piece of information and thereby enables the full characterization of the temperature evolution of the magnetic structure of NdFeAsO. The SDW ordering of the Fe moments in NdFeAsO occurs at 135 K, and is characterised by a magnetic propagation vector **k** = (1 0 ½) and a neutron moment of 0.25(5) $\mu_B$ [18]. Upon ordering of the Nd sublattice, the magnetic propagation vector changes to **k** = (1 0 0) and the Fe moment increases to 0.9(1) $\mu_B$ [7]. The Nd moment is 1.55(4) $\mu_B$. The enhancement of the Fe moment was subsequently attributed to the suppression of magnetic fluctuations in the SDW state [19]. Initially, a single Nd ordering at ~ 2 K was reported [7], but a recent single crystal study reveals a more complex sequence of transitions [20]. Upon cooling, a small contribution to the magnetic $(100)_m$ reflection first develops below 15 K, followed by long range Nd ordering at $T_N$ = 6 K. Below 6 K, the intensity of the $(100)_m$ reflection increases linearly and does not saturate. We found a similar temperature dependence with a small Nd moment persisting above 2 K for polycrystalline NdFeAsO using inelastic neutron scattering, revealing that there is no difference between polycrystalline and single crystal samples [21].

A 5 gram polycrystalline sample was prepared using standard solid state chemistry methods following the procedure described in Ref. [22]. The neutron powder diffraction experiment was performed on the D20 beam line at the Institute Laue Langevin, Grenoble France [23]. The instrument was used in the high-flux setting with wavelength $\lambda = 2.419$ Å. Data were collected over a period of 6 hours at a temperature well below (1.65(5) K) and above (30.0(1) K) the Nd ordering temperature. The collected data were normalized to a monitor count of 50000 and subtracted to reveal the small magnetic peaks (Fig. 1). Rietveld fits were performed using the GSAS and EXPGUI programs [24]. Only the background and magnetic parameters were allowed to vary in the magnetic refinements. No negative magnetic intensities indicative of the SDW state were observed in the difference pattern. This is consistent with the small magnitude of the ordered magnetic moment (0.3 $\mu_B$), which is at the sensitivity limit of the D20 instrument. At 1.65(5) K, the Nd and Fe sublattices contribute to the same reflections, and well defined peaks are observed. The different Q-dependence of the magnetic form factors means that by fitting the whole pattern the Nd and Fe contributions can be accurately separated. The histogram scale factor was determined by a fit to the 30 K dataset. This revealed $Nd_2O_3$ (2 wt%), $Nd(OH)_3$ (4 wt%), $Fe_2As$ (2 wt%) and FeAs (1 wt%) impurities ($wR_p = 2.5\%$, $R_p = 1.8\%$ and $R_F^2 = 1.8\%$). The symmetry analysis was done using version 2 K of the Sarah representational analysis program [13]. In agreement with previous results, the magnetic reflections are indexed on the nuclear cell, and the magnetic propagation vector **k** = (1 0 0). The Cmma space group has 2 centring operations and 8 symmetry operations. The 8 symmetry operations leave the propagation vector invariant or transform it into an equivalent vector, and form the little group $G_\mathbf{k}$. The decomposition of the magnetic representation, $\Gamma_{Mag}$ in terms of the non-zero irreducible representations (IRs) of $G_\mathbf{k}$ is $\Gamma_{Mag} = \Gamma_1 + \Gamma_2 + \Gamma_3 + \Gamma_4 + \Gamma_5 + \Gamma_6$ for the Fe site, and $\Gamma_{Mag} = \Gamma_2 + \Gamma_3 + \Gamma_5 + \Gamma_6 + \Gamma_7 + \Gamma_8$ for the Nd site. In case of a second order phase transition, Landau theory states that a single IR becomes critical. The basis vectors describing the combined Nd and Fe magnetic order are therefore associated with one of the $\Gamma_2$, $\Gamma_3$, $\Gamma_5$ or $\Gamma_6$ IRs. The $\Gamma_3$ and $\Gamma_5$ models are ferromagnetic (FM) and can therefore be discarded, leaving the antiferromagnetic (AF) Fe-$\Gamma_2$; Nd-$\Gamma_2$ and Fe-$\Gamma_6$; Nd-$\Gamma_6$ models (Table I). Interestingly, both solutions have the Nd and Fe sublattice magnetization directions arranged perpendicular. This is the same result as derived by Maeter et al. in Ref. [4]. Two further models with both the Nd and Fe moments pointing along the *a*-axis direction (Fe-$\Gamma_2$; Nd-$\Gamma_6$) or *c*-axis direction (Fe-$\Gamma_6$; Nd-$\Gamma_2$) were also tested. The refined moments and fit statistics for these models with a single magnetic variable per sublattice are given in Table 1. The Rietveld fits are shown in Fig. 1. The Fe-$\Gamma_2$; Nd-$\Gamma_2$ model has an unchanged SDW iron stripe ordering with Fe moments AF coupled along the "long" *a*-axis and FM along the "short" *b*-axis, and moments aligned along the *a*-direction. The Nd moments are perpendicular to the basal plane. This solution underestimates the intensity of the (100) and (103) reflections, and overestimates the (210) reflection. The Fe-$\Gamma_6$; Nd-$\Gamma_6$ model maintains the Fe spin-stripe but now with moments aligned along the *c*-axis. The Nd moments are in the basal plane and point along the *a*-axis. This improves the fit, and the (100), (103) and (201) reflections are now properly taken into account. However, this solution slightly overestimates the weak (102) reflection and somewhat underestimates the (211) reflection. The Fe-$\Gamma_2$; Nd-$\Gamma_6$ and Fe-$\Gamma_6$; Nd-$\Gamma_2$ models do not fit the data as well, revealing that both $m_x$ and $m_z$ components are needed to adequately fit the data. The absolute values of the fit statistics have to be treated with some caution as the difference pattern was offset to eliminate negative intensities. However, the observed trends remain valid. A LeBail fit yielded $\chi^2 = 5$ and a background subtracted $wR_p = 6.3\%$. The best fit is obtained with the Fe-$\Gamma_6$; Nd-$\Gamma_6$ solution, which has $\chi^2 = 9.7$ and a background subtracted $wR_p = 8.1\%$. This is substantially better the Fe-$\Gamma_2$; Nd-$\Gamma_2$ solution, which has 15-20% higher $\chi^2$ and $wR_p$ values. The refined moments for these models are $m_{Fe} = 0.5$ $\mu_B$ and $m_{Nd} = 0.9$ $\mu_B$ (Table I).

A further symmetry allowed magnetic model can be constructed by taking a linear combination of the $\Gamma_2$ and $\Gamma_6$ solutions. This doubles the number of magnetic variables, which makes it difficult to determine whether any improvement in the fit is statistically significant. The Fe-($\Gamma_2 \times \Gamma_6$); Nd-($\Gamma_2 \times \Gamma_6$) linear combination fits the data well and yields the following fit statistics: $\chi^2 = 6.2$, $wR_p = 6.3\%$ and $R_F^2 = 13.1\%$. The refined moments are $m_{Fe,x} = 0.35(5)$ $\mu_B$, $m_{Fe,z} = 0.34(4)$ $\mu_B$, $m_{Fe} = 0.48(3)$ $\mu_B$; $m_{Nd,x} = 0.73(3)$ $\mu_B$, $m_{Nd,z} = 0.53(5)$ $\mu_B$, $m_{Nd} = 0.90(2)$ $\mu_B$. This solution has the Nd and Fe sublattice magnetizations parallel and maintains the same Fe and Nd moments as found for the models with two magnetic variables. The previously reported model [3 magnetic variables [7], in our notation Fe-$\Gamma_2$; Nd-($\Gamma_2 \times \Gamma_6$)] also fits the data well and yields $m_{Fe,x} = 0.53(2)$ $\mu_B$, $m_{Nd,x} = 0.56(2)$ $\mu_B$ and $m_{Nd,z} = 0.73(2)$ $\mu_B$, $m_{Nd} = 0.92(2)$ $\mu_B$ ($\chi^2 = 7.7$, $wR_p = 7.0\%$ and $R_F^2 = 14.7\%$). The same Fe and Nd moments are therefore also found for this solution. There is however no symmetry reason to constrain the Fe moment to lie along the a-axis, while allowing two magnetic variables for the Nd sublattice. Any attempt to include a magnetic $m_y$ component on the Nd sublattice results in unstable refinements, which appears to exclude a non-collinear Nd ordering. In addition, there is no evidence for the existence of more than one **k** vector [4].

The main results from the combined magnetic symmetry and neutron powder diffraction analysis can be summarized as follows: (a) The Fe-$\Gamma_6$; Nd-$\Gamma_6$ model (shown in Fig. 2) with two magnetic variables fits the data well, and is the simplest symmetry derived model to do so. For this reason, it is preferred, but the models with more magnetic variables cannot be totally excluded based on the present data. This result suggests that the Fe moments undergo a spin-reorientation upon magnetic ordering of the Nd sublattice, and that the Fe and Nd moments have a perpendicular orientation. This is consistent with the observation that isotropic antiferromagnetic exchange between the R/TM sublattices is frustrated (Fig. 2 and [4, 6, 15]). A similar transition is observed in NdMnAsO, where the orientation of the Mn moment changes from parallel to the *c*-axis into the *ab*-plane, while the Nd moment lies in the basal plane [15, 16]. For NdCoAsO, both the Nd and Co moments are found to lie in the basal plane [14, 17]. This orientation of the Nd moments is likely to be due to the crystal field splitting of the *4f* states, which is expected to be similar for the NdTMAsO materials. (b) the Fe (0.5 $\mu_B$) and Nd (0.9 $\mu_B$) moments are the same in all solutions that fit the data well, and thus irrespective of the orientation of the sublattice magnetization. These values are smaller than observed by Qiu et al. [$m_{Fe}$ = 0.9(1) $\mu_B$; $m_{Nd}$ = 1.55(4) $\mu_B$] and in our earlier work [$m_{Fe}$ = 1.1(1) $\mu_B$; $m_{Nd}$ = 1.90(3) $\mu_B$] [7, 22]. The discrepancy results from the correlation between the phase scale factor and the weight fractions of the nuclear impurity phases. Partially, or not taking into account the impurities, results in an inflated scale factor, and artificially enhanced magnetic moments. For the current data, the refined magnetic moments are doubled ($m_{Fe}$ = 1.1(1) $\mu_B$ and $m_{Nd}$ = 1.9(1) $\mu_B$) if the impurity phases are not taken into account. The subtraction of the 30 K and 1.6 K data enables the separation of the magnetic from the nuclear contributions, and eliminates any effects resulting from peak overlap. This affords the reliable determination of the direction of the moments, and their relative values. However, the *magnitude* of the moments is determined by the phase scale factor, and this causes the largest errors in the reported moments from neutron powder diffraction. Note that the magnetic models in Table 1 yield similar moments even when they do not fit the data well. We recently used inelastic neutron scattering to directly probe the ordered Nd moment [21]. This yielded a moment of 1.0 $\mu_B$ at 1.65 K, which is in excellent agreement with the value obtained here. Upon further cooling, the Nd moment increases gradually to 1.25 $\mu_B$ at 55 mK. This induced temperature dependence is similar to that observed for the Nd moments in NdMnAsO and NdCoAsO [14, 15, 17]. In these two materials the transition metal moment does not change upon Nd ordering, which strongly suggests that the Fe moment in NdFeAsO remains unchanged upon further cooling, and that our value of 0.5 $\mu_B$/Fe is reliable. Finally, heat capacity measurements reveal a Rln(2) magnetic entropy contribution associated with the Nd ordering [25], which is consistent with the Nd moment observed here.

To conclude: the low-temperature magnetic structure of NdFeAsO has been reinvestigated using neutron powder diffraction and symmetry analysis. The best fit is obtained using a model with Nd moments aligned along the a-axis, and Fe moments along the c-axis. The fitted moments are $m_{Nd}$ = 0.9 $\mu_B$ and $m_{Fe}$ = 0.5 $\mu_B$. This brings the low-temperature Fe moment in line with the values observed in the SDW state, and reveals there is no substantial enhancement upon Nd ordering.

**Acknowledgements**

J-WGB acknowledges the ILL for discretionary beam time, and the Royal Society and EPSRC-GB for support.

**Figure Captions**

Fig. 1. Rietveld fits (solid red line) to the magnetic 1.65(5) – 30.0 K difference pattern (open circles). Magnetic Bragg reflection markers and difference curves are shown. The data are offset by 20 counts to eliminate negative intensities. The symmetry labels of the fitted models correspond to the ones shown in Table 1. Neutron diffraction data were collected over 6 hours for each pattern and normalized to a monitor count of 50000. The (101) reflection has 150 counts, which is equivalent to 3.5% of the most intense nuclear reflection. The S-shaped features are due to the lattice expansion of NdFeAsO.

Fig.2. Schematic representation of the Fe-$\Gamma_6$; Nd-$\Gamma_6$ model. Nd: large yellow spheres, Fe: small blue spheres. The lines are guides to the eye indicating the topology of the magnetic interactions.

Table I. Basis vectors [$m_x$, $m_y$, $m_z$] for space group Cmma with magnetic propagation vector **k** = (1 0 0), refined Fe and Nd moments and goodness of fit statistics.

| | | Fe-$\Gamma_2$; Nd-$\Gamma_2$ | Fe-$\Gamma_2$; Nd-$\Gamma_6$ | Fe-$\Gamma_6$; Nd-$\Gamma_2$ | Fe-$\Gamma_6$; Nd-$\Gamma_6$ |
|---|---|---|---|---|---|
| | Fe1 | [$m_x$, 0, 0] | [$m_x$, 0, 0] | [0, 0, $m_z$] | [0, 0, $m_z$] |
| | Fe2 | [-$m_x$, 0, 0] | [-$m_x$, 0, 0] | [0, 0, –$m_z$] | [0, 0, –$m_z$] |
| | Nd1 | [0, 0, $m_z$] | [$m_x$, 0, 0] | [0, 0, $m_z$] | [$m_x$, 0, 0] |
| | Nd2 | [0, 0, –$m_z$] | [-$m_x$, 0, 0] | [0, 0, –$m_z$] | [-$m_x$, 0, 0] |
| Fe1 | $m_x$ ($\mu_B$) | 0.45(3) | 0.45 | 0 | 0 |
| | $m_z$ ($\mu_B$) | 0 | 0 | 0.40(10) | 0.53(2) |
| Nd1 | $m_x$ ($\mu_B$) | 0 | 1.07(5) | 0 | 0.88(2) |
| | $m_z$ ($\mu_B$) | 0.90(1) | 0 | 1.03(4) | 0 |
| | $\chi^2$ | 11.3 | 54.7 | 28.7 | 9.7 |
| | w$R_p$ (%) | 10.0 | 24.1 | 15.5 | 8.1 |
| | $R_F^2$ (%) | 18.1 | 19.6 | 24.6 | 16.8 |

Fe1 (0.25, 0, 0.5); Fe1 (075, 0, 0.5); Nd1 (0, 25, 0.14); Nd2 (0, 0.75, 0.86).

Fig. 1

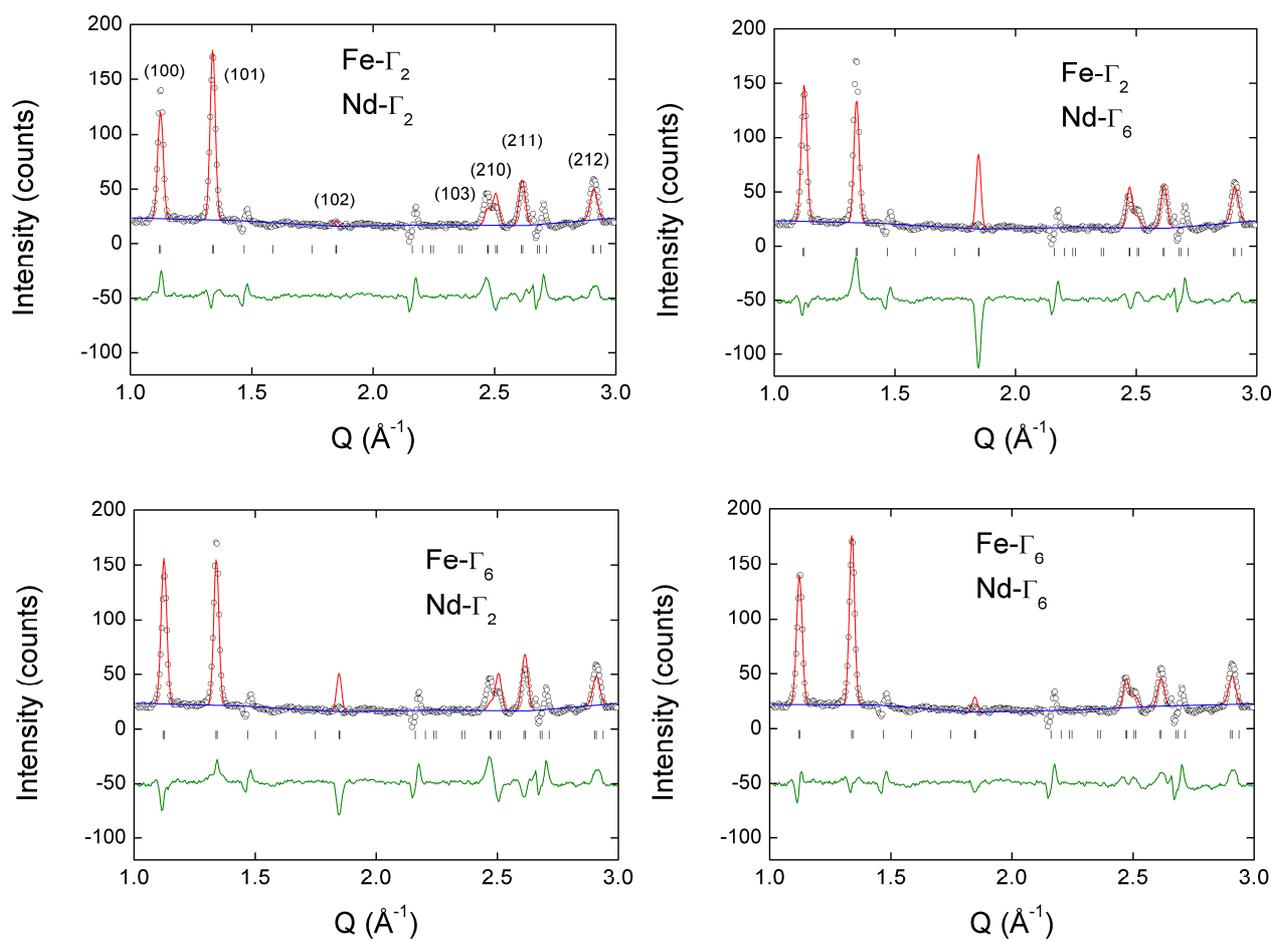

Fig. 2.

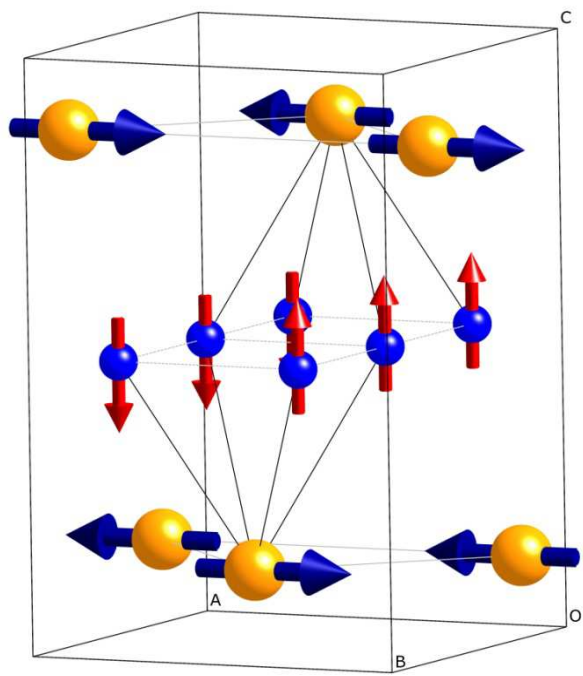